\documentclass
[aps,pra,amsfonts,amssymb,twocolumn,amsmath,preprintnumbers,nofootinbib,floatfix,
showpacs]{revtex4-1}%
\usepackage[dvips]{graphics}
\usepackage{graphicx}
\usepackage{dcolumn}
\usepackage{bm}
\usepackage{amsmath}
\usepackage{amsfonts}
\usepackage{amssymb}
\usepackage{xcolor}
\usepackage{subfigure}
\usepackage{hyperref,hypcap}
\usepackage{braket}
\usepackage{commath}%
\providecommand{\U}[1]{\protect\rule{.1in}{.1in}}
\begin{document}

\title{Topological electric driving of magnetization dynamics in insulators}
\author{Cong Xiao}
\thanks{These authors contributed equally to this work.}
\author{Bangguo Xiong}
\thanks{These authors contributed equally to this work.}
\author{Qian Niu}
\affiliation{Department of Physics, The University of Texas at Austin, Austin, Texas 78712, USA}

\begin{abstract}
Established forms of electromagnetic coupling are usually conservative
(in insulators) or dissipative (in metals and semiconductors). Here we point out
the possibility of nondissipative electric driving of magnetization dynamics,
if the valence electronic states have nontrivial topology in the combined space of
crystal momentum and magnetization configuration. We provide a hybrid insulator system
to demonstrate that the topology-based nonconservative electrical generalized
force is capable of supporting sustained magnetization motion in the presence of
Gilbert damping, with quantized and steady energy pumping into magnetization motion
from the electric field. We also generalize our results to magnetic textures,
and discuss electric field induced Dzyaloshinskii-Moriya interaction which can be nonconservative.
\end{abstract}
\maketitle

%\author{Cong Xiao}
%\thanks{These authors contributed equally to this work.}
%\author{Bangguo Xiong}
%\thanks{These authors contributed equally to this work.}
%\author{Qian Niu}
%\affiliation{Department of Physics, The University of Texas at Austin, Austin, Texas 78712, USA}

\section{Introduction}

The study of electrical control of magnetization dynamics has occupied a large
part of solid state research for many decades, which generally falls into two
separate categories known as multiferroics
\cite{Nagaosa2005,Maxim2006,Tokura2007} and spintronics \cite{Zutic2004}
depending on the conductive behavior of the hosting materials. The former
deals with insulators where electrical effects on magnetization is
characterized through the free energy \cite{Han2013}, and the resulting torque
would be naturally considered as conservative and unable to drive sustained
motion of the magnetization for a static electric field. In the latter, one
finds various current induced magnetic torques in metals and semiconductors
\cite{Manchon2008,Garate2009,Kurebayashi2014}, which can provide a persistent
source of energy for sustained motion of magnetization, but one has to deal
with wasteful and prohibitive joule heating in practice.

Magnetic insulators have recently been utilized to achieve low-dissipation
magnetization control by combining the insulator with heavy metals hosting
prominent spin Hall effect that injects a spin current into the insulator
\cite{Beach2017,Shi2018}. An electric field can also directly manipulate
magnetization in an insulator without Joule heating by means of spin-orbit
torques mediated by occupied electronic states
\cite{Garate2010,Nagaosa2011,Loss2012}. In particular, mesoscopic transport
theories proposed the exchange gapped edge states of a two
dimensional topological insulator in hybrid with a magnet as a unique platform for
studying the magnetic Thouless motor \cite{Felix2013,Felix2015}, which works as
the inverse mode of the adiabatic charge pumping by a cyclic magnetic motion
\cite{Qi2008-1,Thouless1983,Meng2014} under an applied voltage. By using the
scattering matrix approach \cite{Felix2011}, previous works showed quantized
electrical energy transfer into the magnet if the magnetization accomplishes a
cyclic motion \cite{Felix2013,Felix2015}. On the other hand, as the Berry curvature
in the mixed space of crystal momentum and magnetization configuration underlies
the magnetic Thouless pumping, it would be interesting to reveal the relation
between nonzero electrical energy input into magnetic dynamics and the
topological characteristics in the mixed parameter space \cite{Yuriy2017}.
Moreover, it has not been shown whether the electric driving of sustained
magnetic motion, which enables a motor, can be realized in the presence of
magnetic damping due to coupling of magnetization to other degrees of freedom
than electrons.

In this study we show the possibility of nondissipative driving of
magnetization dynamics with steady energy pumping by a static electric field
in insulators. This is motivated by the fact that electric polarization is not
always a single-valued quantity \cite{KS1993,Nagaosa2004}, and the adiabatic
current pumped by cyclic motion of the magnetization can acquire a quantized
net amount of energy from a static electric field under certain topological
conditions of the valence electronic states. We exploit this idea in a model
system of edge states of a two-dimensional topological insulator gapped by
hybrid with a magnetic wire, and show explicitly sustained magnetization
motion when a constant electric field is applied to overcome Gilbert damping.

Our results can also be generalized to the case of slowly varying magnetic
textures. There is a topological current bilinear in the gradient and time
derivative of the magnetization density \cite{Xiao2009}, a sort of anomalous
Hall current induced by the artificial electric field from the time dependent
magnetic texture \cite{Yaroslav2008,Volovik1987,Duine2008,Yang2009}. This
current provides a channel of nondissipative drive of the magnetic texture by
an external static electric field. In topologically nontrivial cases this
drive is nonconservative and capable of delivering a nonzero and quantized
amount of energy when the magnetic texture wraps around the Bloch sphere in
time. In topologically trivial cases, where the electric polarization induced
by magnetization gradient is well defined, the drive is conservative because
it can be identified as originating from a polarization energy density whose
susceptibility to the magnetization gradient gives the electric field induced
Dzyaloshinskii-Moriya interaction (DMI)
\cite{DM,DM-1,Freimuth2013,Freimuth2014}.

The rest of the paper is organized as follows. In Sec. II we focus on the
electric-field induced generalized force on a homogeneous magnetization in
insulators, and study its nonconservative nature which is related to certain
topological conditions of the occupied electronic states. These general
rationales are illustrated in a hybrid insulator in Sec. III, in which the
electric driving of sustained magnetic motion is also demonstrated. Section IV
is devoted to the electrical generalized force on the magnetization in
inhomogeneous insulators and its relation to an electric-field induced DMI.
Finally, we concludes the paper in Sec. V.

\section{Electrical generalized force on magnetization}

In the language of analytic mechanics, an generalized force is an amount of
work done on the system per unit displacement in the dynamical variable (the
magnetization here). Considering a system with a homogeneous magnetization
$\boldsymbol{m}$ coupled to an electronic insulator, change in the
magnetization can in general pump an adiabatic electric current
\begin{equation}
\boldsymbol{j}=e\int\left[  d\boldsymbol{k}\right]  \Omega_{\boldsymbol{km}%
}\cdot\dot{\boldsymbol{m}},
\end{equation}
where $\Omega_{\boldsymbol{km}}$ is the electronic Bloch-state Berry curvature
in the parameter space of the magnetization and crystal momentum
$\boldsymbol{k}$ (set $\hbar=1$ unless otherwise noted), with its Cartesian
components given by $-2\mathrm{Im}\langle{\partial}_{k{_{i}}}{u}|{\partial
}_{m{_{j}}}{u}\rangle$. Here $|u\rangle$ is the periodic part of the Bloch
wave, and the band index $n$ is omitted for simple notation. $\left[
d\boldsymbol{k}\right]  \equiv d^{d}k/(2\pi)^{d}$ with $d$ as the spatial
dimension, and the summation over valence bands is implied. Through this
adiabatic current, an external electric field can deliver work on the system,
with the work density $\delta w=\boldsymbol{E}\cdot\boldsymbol{j}dt$, which is
proportional to $\delta\boldsymbol{m}=\dot{\boldsymbol{m}}\cdot dt$.
Therefore, we obtain the electrical generalized force density on magnetization
as%
\begin{equation}
\boldsymbol{\mathcal{\boldsymbol{E}}}_{m}^{e}\equiv\frac{\delta w}%
{\delta\boldsymbol{m}}=e\boldsymbol{E}\cdot\int\left[  d\boldsymbol{k}\right]
\Omega_{\boldsymbol{km}}. \label{field}%
\end{equation}

This electrical generalized force is nondissipative because of the lack of
conduction electrons for joule heating, and is in fact topological in the
sense that it delivers a quantized amount of energy over a cycle of the
magnetization motion. For simplicity, we first consider an insulator with zero
Chern numbers in the Brillouin zone at each point over the path of
$\boldsymbol{m}$, such that one can take a $\boldsymbol{k}$-space periodic
gauge to locally define an electronic polarization $\boldsymbol{P}%
=-e\int\left[  d\boldsymbol{k}\right]  \boldsymbol{\mathcal{A}}%
_{\boldsymbol{k}}$ \cite{KS1993}, with $\boldsymbol{\mathcal{A}}%
_{\boldsymbol{k}}=\langle{u}|i\partial_{\boldsymbol{k}}{u}\rangle$. Then the
electrical work density delivered over the cycle can be written in terms of
the change of this polarization
\begin{equation}
w=\oint d\boldsymbol{m}\cdot\boldsymbol{\mathcal{\boldsymbol{E}}}_{m}%
^{e}=\boldsymbol{E}\cdot\Delta\boldsymbol{P}. \label{W}%
\end{equation}
This change is quantized in units of $\Delta\boldsymbol{P}=-e\boldsymbol{a}%
/V_{0}$ with $\boldsymbol{a}$ being a discrete lattice vector (including the
null vector) and $V_{0}$ the volume of a unit cell. When this change is zero,
so that the polarization is globally defined, the electrical generalized force
is conservative in the sense that its work can be regarded as a change in the
globally well defined polarization energy density $-\boldsymbol{E}%
\cdot\boldsymbol{P}$. When this change is nonzero, the electrical generalized
force is nonconservative and capable of supporting sustained magnetization
motion even in the presence of Gilbert damping due to other dissipative channels.

Some comments are in order. First, if the electronic insulator is one
dimensional, then the electrical work (per unit length) over a cycle of the
magnetization reduces to $eE$ times the Chern number over the torus of the
combined space of crystal momentum and the magnetization (along its path),
corresponding to the quantized number of electrons pumped over the cycle.
Second, quantization of electrical work over the cycle of magnetization also
applies to insulators with nonzero $\boldsymbol{k}$-space Chern numbers by a
simple argument. Although one cannot take a periodic gauge in $\boldsymbol{k}%
$-space, one can always choose a periodic gauge over a fixed one dimensional
path of the magnetization. It is then clear that the electrical work over the
cycle equals the Brillouin-zone integral of the $\boldsymbol{k}$-gradient of
the Berry's phase over the cycle. Topological quantization of this work then
follows from the multi-valuedness of the Berry's phase. Third, using the
Bianchi identity on Berry curvatures, one can easily show that the electrical
generalized force is curl-free $\partial_{\boldsymbol{m}}\times
\boldsymbol{\mathcal{\boldsymbol{E}}}_{m}^{e}=0$ everywhere in $\boldsymbol{m}%
$-space, except the singular points where the energy gap above the filled
states of the electron system closes.

\begin{figure}[ptb]
\includegraphics[width=1\columnwidth]{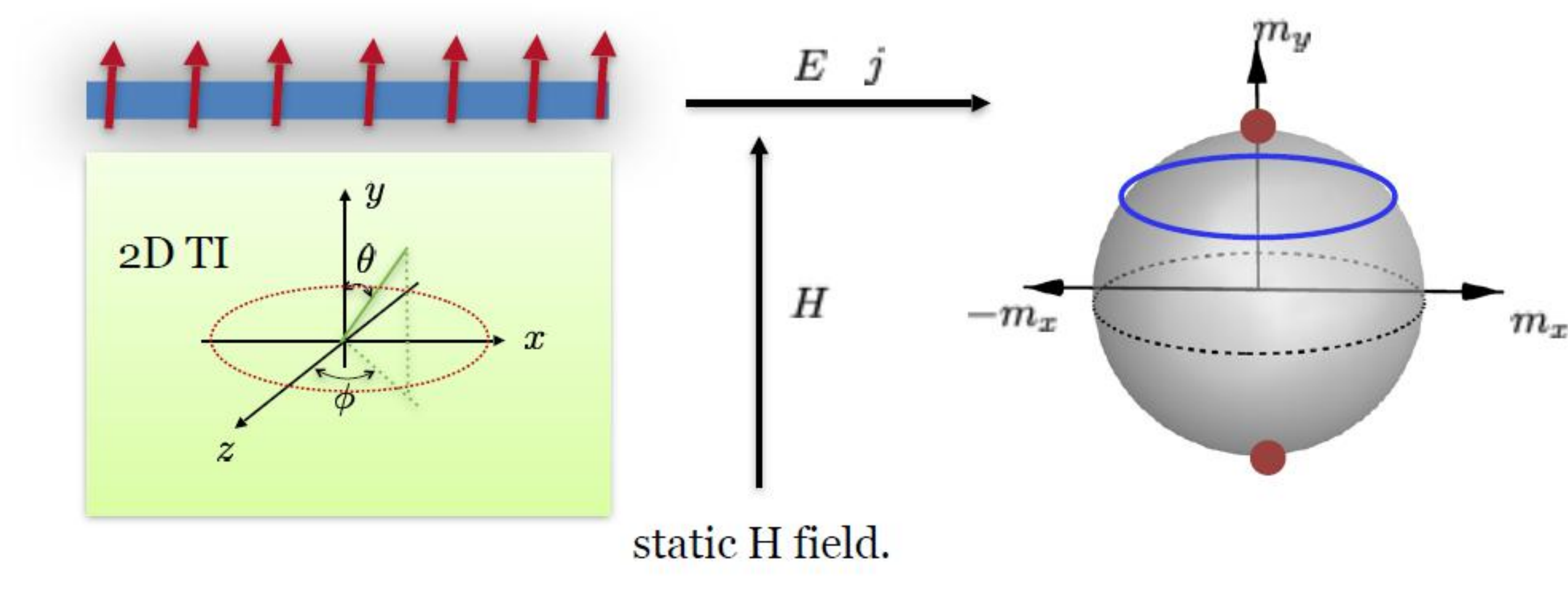}\caption{A ferromagnetic wire
(blue bar) hybridizes and gaps the edge states of a two-dimensional
topological insulator (green region). When the magnetization $\bm{m}$ moves
around (blue circle on the right) on the Bloch sphere, the pumped adiabatic
current $j$ along the edge couples to an applied electric field $\bm{E}$ to
provide energy to overcome Gilbert damping. A static magnetic field $\bm{H}$
is applied to help preparing the system into a sustained motion of limit
cycle. }%
\label{fig:setup}%
\end{figure}

When can the electrical work on magnetization be nonzero? Quantization of its
value implies that the electrical work is invariant if the path in
$\boldsymbol{m}$-space is deformed without closing the energy gap. In
particular, within a singly connected region where the gap is open, the
electrical work is zero on all closed paths. This applies for example to the
north or south hemispheres of magnetization in the two dimensional
ferromagnetic Dirac model studied in \cite{Xiong2018}, where one can define
polarization energies separately for each region, although cannot globally
because of gap closing on the equator. Consequently, this model system cannot
provide nonzero electrical work for sustained magnetization motion. It is
therefore clear that the singular points of gap closing have to be arranged to
define multiply connected regular regions, where electrical work can possibly
be nonzero on topologically nontrivial paths.

\section{A model for sustained magnetization motion}

Here we propose a one dimensional model system, where the gap closes on the
two poles of the magnetization Bloch sphere, and the electrical work per unit
length is $eE$ times the winding number of the path around the poles. The
system is constructed by interfacing a magnetic wire with the topological edge
states of a two-dimensional topological insulator (Fig. \ref{fig:setup}). The
exchange coupling renders the electronic system insulating by opening a gap in
the Dirac spectrum. The relevant low-energy Hamiltonian is
\begin{equation}
\hat{h}=\hbar vk\hat{\sigma}_{y}+J\hat{\boldsymbol{\sigma}}\cdot
\boldsymbol{m},\label{model}%
\end{equation}
where $v$ is the Fermi velocity, $\hat{\boldsymbol{\sigma}}$ is the Pauli
matrix, and $J$ is the coupling constant. The magnetization $\boldsymbol{m}$
is assumed to have a fixed magnitude and is parameterized by the polar angle
$\theta$ relative to the $y$ axis and the azimuthal angle $\phi$ as shown in
Fig. \ref{fig:setup}. The energy gap is open everywhere except at the north
and south poles of the Bloch sphere with $m_{y}=\pm m$ (red dots). Assuming
that the lower band is filled and the electric field is applied along the
magnetic wire (positive $x$ direction), we can evaluate the formula for the
electrical generalized force to find
\begin{equation}
\boldsymbol{\mathcal{\boldsymbol{E}}}^{e}=\frac{-eE}{2\pi m}\frac
{\boldsymbol{\hat{e}}_{\phi}}{\sin\theta}=-eE\frac{\partial_{\boldsymbol{m}%
}\phi}{2\pi}.\label{force}%
\end{equation}
It is singular at the poles and is a gradient of the multiple-valued azimuthal
angle, so that the electrical work density over a closed path on the Bloch
sphere is quantized in terms of the winding number of the path
\begin{equation}
\oint d\boldsymbol{m}\cdot\boldsymbol{\mathcal{\boldsymbol{E}}}^{e}%
=-N_{t}eE,\label{work}%
\end{equation}
in line with the aforementioned general topological arguments. The winding
number $N_{t}$ counts how many times the closed path wraps around the $y$ axis counter-clockwise.

In such a one dimensional insulator it is also interesting to understand the
electrical generalized force from the polarization as
$\boldsymbol{\mathcal{\boldsymbol{E}}}^{e}=E\partial_{\boldsymbol{m}}P$, where
the polarization is not single-valued and can only be determined to be
$P=-e\phi/2\pi$ up to an uncertainty quantum $-e$. Consistently, the two gap
closing poles are singular points of the polarization, and the change of
polarization upon a closed path on the Bloch sphere is $-eN_{t}$.

We now proceed to study the dynamics of the magnetization to see the effect of
this generalized force. In the absence of coupling to the electronic system,
we can rewrite the Landau-Lifshitz-Gilbert equation of the ferromagnet in the
form of $-\partial_{\boldsymbol{m}}\mathcal{G}^{0}+\dot{\boldsymbol{m}}%
\times\boldsymbol{\Omega}_{\boldsymbol{m}}^{0}-\eta^{0}\dot{\boldsymbol{m}}%
=0$, as balancing out a conserved force from the free energy $\mathcal{G}^{0}%
$, a Lorentz type force from the $\boldsymbol{m}$-space Berry curvature
$\boldsymbol{\Omega}_{\boldsymbol{m}}^{0}$ \cite{Niu1999}, and a frictional
force with a scalar damping coefficient $\eta^{0}$. We will model the free
energy density as $\mathcal{G}^{0}=-K^{0}\hat{m}_{x}^{2}-Hm_{y}$ with an easy
axis anisotropy and an applied static magnetic field $H$. The $\boldsymbol{m}%
$-space Berry curvature is given in terms of the gyromagnetic ratio
$\gamma^{0}$ as $\boldsymbol{\Omega}_{\boldsymbol{m}}^{0}=\boldsymbol{m}%
/(m^{2}\hbar\gamma^{0})$. The damping coefficient is related to the Gilbert
number $\lambda$ as $\lambda=(\gamma^{0})^{2}\eta^{0}$.

\begin{figure}[ptb]
\includegraphics[width=1\columnwidth]{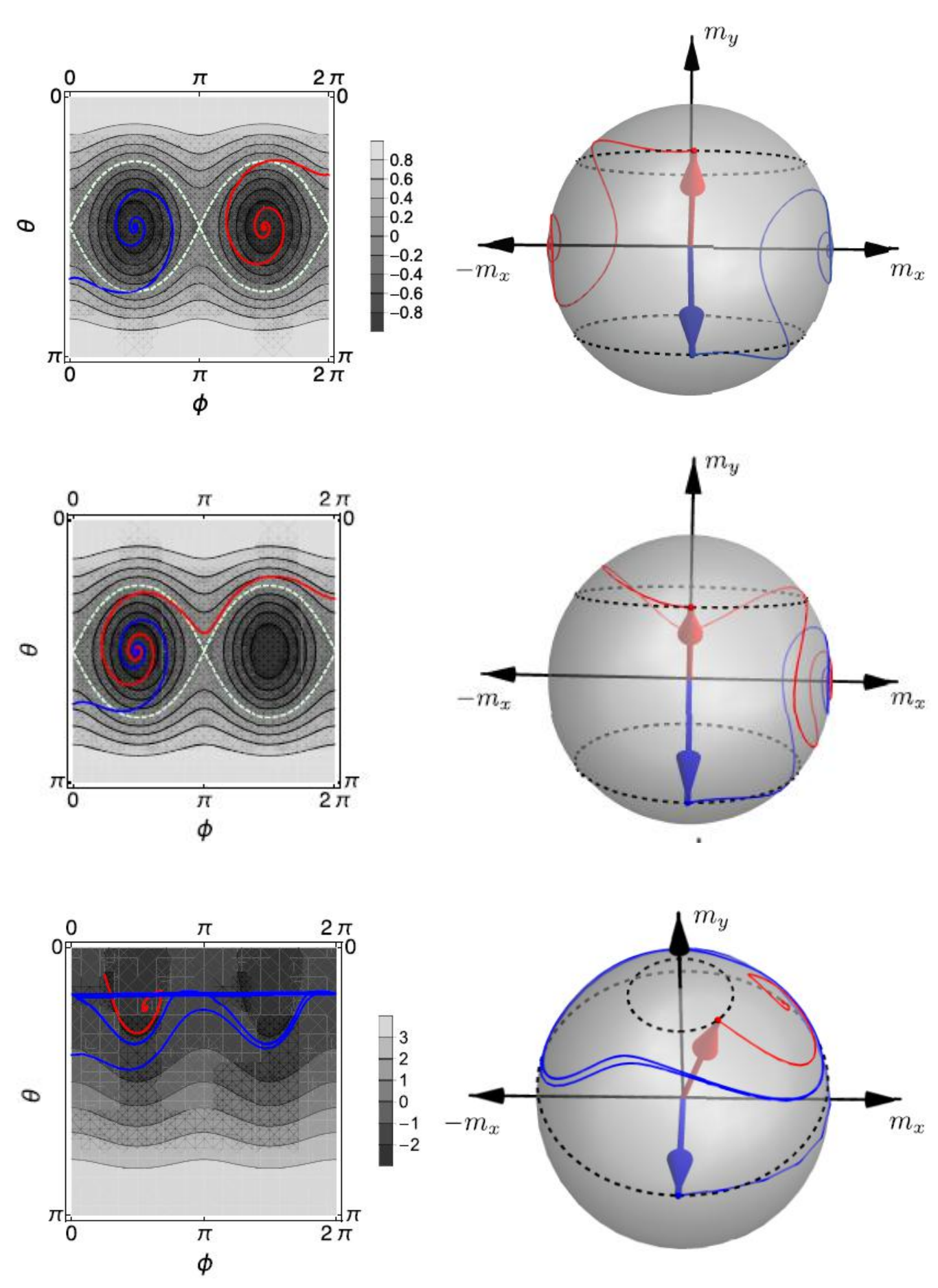}\caption{ Free energy
contours in the angular space and typical evolution trajectories on the Bloch
sphere in the absence (top panels) and presence (middle panels) of an electric
field, and in the presence of both electric and magnetic fields (bottom
panels). In the last case, a limit cycle emerges.}%
\label{fig:evolution}%
\end{figure}

In the presence of coupling to the electronic system, the equation of motion
becomes%
\begin{equation}
\boldsymbol{\mathcal{\boldsymbol{E}}}_{m}^{e}-\partial_{\boldsymbol{m}%
}\mathcal{G}+\dot{\boldsymbol{m}}\times\boldsymbol{\Omega}_{\boldsymbol{m}%
}-\eta\dot{\boldsymbol{m}}=0,
\end{equation}
where the electrical generalized force enters as an extra term along with
electronic modifications to the other terms. The gap opening in the electronic
system contributes a lowering of the free energy $\mathcal{G}^{e}=K^{e}%
(\hat{m}_{y}^{2}-1)$ that we model as a hard-axis anisotropy. The electronic
contribution to the $\boldsymbol{m}$-space Berry curvature\ is given by
$\boldsymbol{\Omega}_{\boldsymbol{m}}^{e}=\int\left[  d\boldsymbol{k}\right]
\Omega_{\boldsymbol{m}}=\boldsymbol{m}/(m^{2}\hbar\gamma^{e})$, where
$\Omega_{\boldsymbol{m}}=\partial_{\boldsymbol{m}}\times
\boldsymbol{\mathcal{A}}_{\boldsymbol{m}}$ is derived from
$\boldsymbol{\mathcal{A}}_{\boldsymbol{m}}=\langle{u}|i\partial
_{\boldsymbol{m}}{u}\rangle$, and $\gamma^{e}=2\pi v/J$. Finally, we assume
that the gap of the electronic system remains open during the course of
dynamics, so there is no electronic contribution to the damping coefficient
$\eta=\eta^{0}$.

Representative results of the magnetization motion are presented in Fig.
\ref{fig:evolution}, where we take $\gamma^{e}/\gamma^{0}=\pi$, $K^{e}%
/(m/\gamma^{0})=K^{0}/(m/\gamma^{0})=1$ GHz and $\eta=0.2/(m\gamma^{0})$.
Shown in the top and middle panels ($H=0$), there are two types of energy
conserved motion in the absence of damping and external fields, divided by the
contour of zero energy (the white dashed curve). In the area enclosing the two
points of lowest energy, $\boldsymbol{m}$ rotates around the $x$ axis, whereas
in the upper and lower areas outside of the zero-energy contour
$\boldsymbol{m}$ goes around the $y$ axis. This situation is changed in the
presence of damping, as shown in the top panel, where two points
$(\phi=0,\theta=0.3\pi)$ and $(\phi=0,\theta=0.7\pi)$ outside of the
zero-energy contour evolve to different points of lowest energy. In the middle
panels, an electric field $eE/2\pi=0.1K^{0}$ is applied, which gives a force
in the clockwise (negative $\phi$) direction. The blue trajectory starting
from $(\phi=0,\theta=0.7\pi)$ falls faster to the $+m_{x}$ axis, while the red
trajectory starting from $(\phi=0,\theta=0.3\pi)$ extends for $3/4$ circle
before the final decay into the same energy minimum as the other trajectory.

\begin{figure}[tbh]
\includegraphics[width=1\columnwidth]{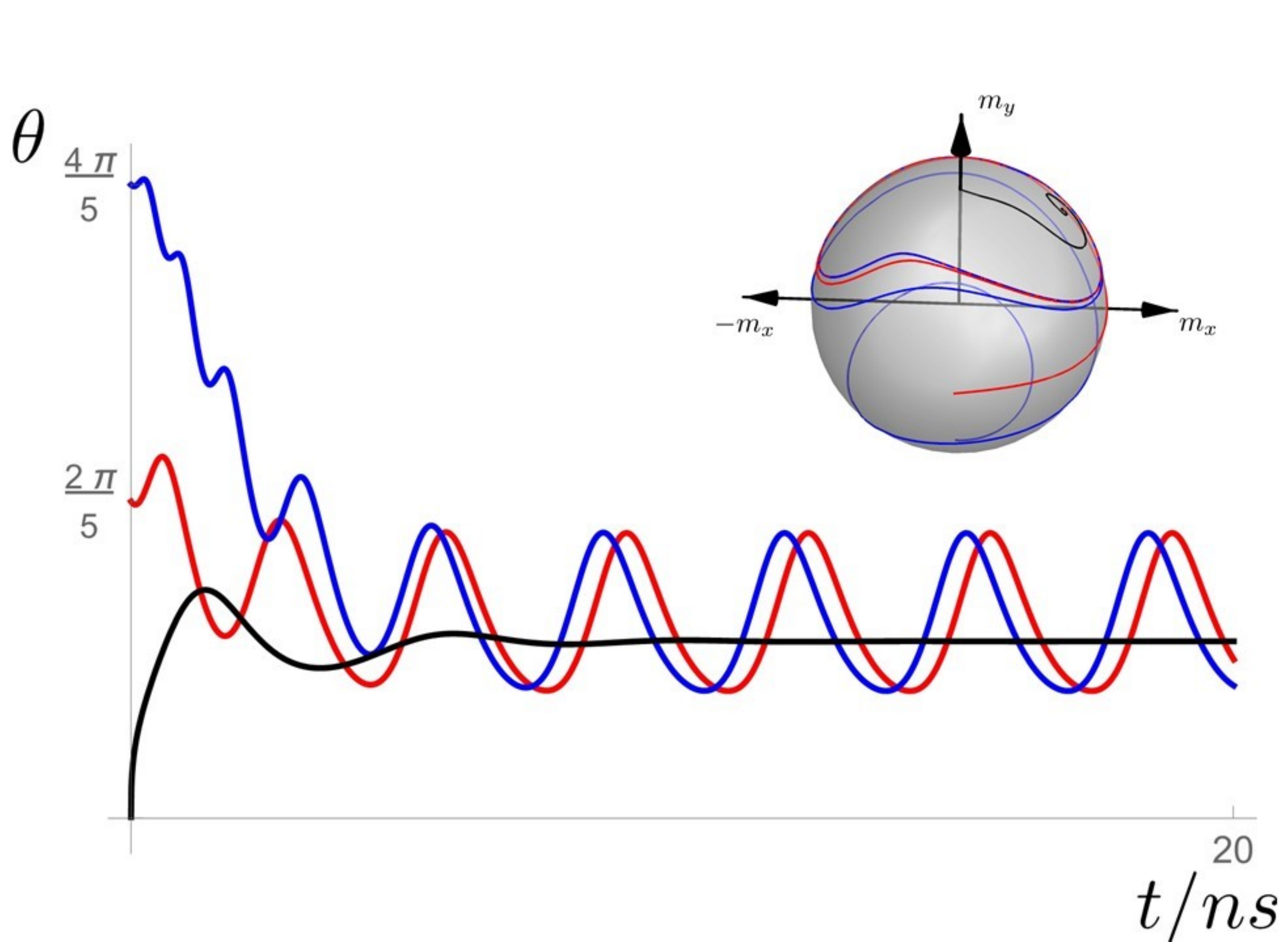}\caption{Time dependence of
the polar angle for different initial conditions, $\phi=0$, $\theta=0.001\pi$
(black), $\theta=0.4\pi$ (red), $\theta=0.8\pi$ (blue). Correspondingly on the
Bloch sphere shown in the inset, the red and blue trajectories evolve into a
right-handed limit cycle, whereas the black trajectory evolves into the point
of lowest energy. }%
\label{fig:angle}%
\end{figure}

The lower panels show the situation where limit cycle motion is found. We
found it important to prepare the system with predominantly around-$m_{y}%
$-axis energy contours, so that the non-conservative electrical force can be
best utilized. We therefore apply a static magnetic field in $y$ direction
with the magnitude $H=K^{0}/m$ to change the energy landscape. We also
switched the direction of the electric field so that the electrical force goes
along the directions of the energy contours. We found that all initial points
in a wide region, between the two dashed circles shown on the right of the
lower panels of Fig. \ref{fig:evolution}, fall into the same limit cycle. For
instance, the blue curve starts from $(\phi=0,\theta=0.4\pi)$ and evolves into
the right handed limit cycle under an electric field $eE/2\pi=-0.1K^{0}$.
Figure \ref{fig:angle} shows how the limit cycle motion is reached in time for
two trajectories (blue and red) from different initial angles, along with one
(black) that falls into an energy minimum. On the limit cycle, we found that
the energy input from the electrical force balances out the energy dissipation
from the Gilbert damping, $\oint d\boldsymbol{m}\cdot
(\boldsymbol{\mathcal{\boldsymbol{E}}}_{m}^{e}-\eta\dot{\boldsymbol{m}})=0$,
as can be easily derived from the equation of motion.

\section{Electrical DMI force}

So far we have been concentrating on nondissipative electrical driving on a
uniform magnetization. When the magnetization is nonuniform, the electrical
generalized force Eq. ({\ref{field}}) still applies as a local force density,
but there will be additional contributions due to the magnetization gradients.
In metals, the electric-current induced DMI have been discussed recently
\cite{Karnad2018,Kato2019,Freimuth2018}, which is similar to the current
induced orbital magnetization \cite{Murakami2015,Mak2017}. The intrinsic
analog, the electric-field induced nondissipative DMI
\cite{Kita1980,Nagaosa2005}, remains elusive in the band picture, but should
be well defined in insulators as we show now.

To first order in the gradient, there is an adiabatic current pumped by the
magnetization dynamics \cite{Xiao2009} $\boldsymbol{j}=e\int\left[
d\boldsymbol{k}\right]  \Omega_{\boldsymbol{k}[\boldsymbol{kr}]\boldsymbol{m}%
}\cdot\dot{\boldsymbol{m}}$ involving the second Chern form of Berry
curvatures $\Omega_{k_{s}[\boldsymbol{kr}]m_{j}}\equiv\Omega_{k_{s}k_{i}%
}\Omega_{r_{i}m_{j}}+\Omega_{k_{s}r_{i}}\Omega_{m_{j}k_{i}}+\Omega_{k_{s}%
m_{j}}\Omega_{k_{i}r_{i}}$. Through this current, an external electric field
can produce a work density $\delta w=\boldsymbol{E}\cdot\boldsymbol{j}dt$
proportional of $\delta m$, implying an electrical generalized force linear in
the magnetization gradient%
\begin{equation}
\boldsymbol{\mathcal{\boldsymbol{E}}}_{m}^{e}=e\boldsymbol{E}\cdot\int\left[
d\boldsymbol{k}\right]  \Omega_{\boldsymbol{k}[\boldsymbol{kr}]\boldsymbol{m}%
.} \label{Chern}%
\end{equation}
For reasons to be discussed later, we will call this an electrical DMI force,
although it is nonconservative in general and capable of sustained driving of
magnetization textures.

Because the second Chern form is antisymmetric in the crystal momentum, a
nonzero result demands that the electronic system is more than one
dimensional. Consider for simplicity a two-dimensional system with the
magnetization gradient in the $y$ direction (one-dimensional domain wall or a
spiral) and an electric field applied in the transverse $x$ direction. The
electrical work per unit transverse width over one pumping period may be
written as
\begin{equation}
W=eE_{x}N_{yt}\int_{T^{2}}\frac{d^{2}k}{2\pi}\int_{S^{2}}\frac{d\theta d\phi
}{2\pi}\Omega_{k_{x}k_{y}\theta\phi}=eE_{x}N_{yt}C_{2}%
\end{equation}
which is topological and quantized in terms of the second Chern number $C_{2}$
in the space \cite{Qi2008} spanned by the Brillouin zone and the Bloch sphere,
and the winding number $N_{yt}=\frac{1}{4\pi}\int dydt\boldsymbol{\hat{m}%
}\cdot(\partial_{y}\boldsymbol{\hat{m}}\times\partial_{t}\boldsymbol{\hat{m}%
})$ for the mapping $\boldsymbol{\hat{m}}(y,t)$ of the $yt$ space-time onto
the Bloch sphere \cite{Braun2012} ($\boldsymbol{\hat{m}}=\boldsymbol{m}/m$).
This winding number has previously appeared in discussion of quantized
electromotive force induced by a moving domain wall \cite{Yang2010}, the so
called ferro-Josephson effect, and the second Chern number may be regarded as
the quantum measure of the anomalous Hall response to this emf \cite{note-j}.
The quantized electrical work is therefore a result of this quantized Hall
current in the direction of the applied electric field.

The same second Chern number has also been introduced in study of electric
charges carried by magnetic textures such as a skyrmion \cite{Freimuth2013},
where it may be understood as the quantum measure of charge response to the
quantized flux of artificial magnetic field \cite{note-j}. This is a sort of
Streda dual effect of the quantum Hall current response to the artificial
electric field of the magnetic texture. This relationship becomes especially
clear in the absence of spin-orbit coupling, where $\Omega_{k_{x}k_{y}%
\theta\phi}=\Omega_{k_{x}k_{y}}\Omega_{\theta\phi}$ and $C_{2}$ reduces to the
first Chern number in $\boldsymbol{k}$-space \cite{Yang2011} which
characterizes the usual quantum anomalous Hall insulators.

In non-Chern insulators where one may choose a periodic gauge in
$\boldsymbol{k}$-space, the electrical generalized force may be written as a
field derivative of the polarization energy,
$\boldsymbol{\mathcal{\boldsymbol{E}}}_{m}^{e}=-\delta_{\boldsymbol{m}}U$,
with \cite{Freimuth2013,Freimuth2014}
\begin{equation}
U=-\int d\boldsymbol{rE}\cdot\boldsymbol{P}=\int d\boldsymbol{r}%
\mathcal{D}_{il}\partial_{i}m_{l},
\end{equation}
where $\boldsymbol{P}$ is the electric polarization induced by magnetization
gradient, including a topological Chern-Simons part \cite{Xiao2009} for which
\begin{equation}
\mathcal{D}_{il}=\frac{e}{2}E_{j}\int\left[  d\boldsymbol{k}\right]
(\mathcal{A}_{k_{j}}\Omega_{k_{i}m_{l}}+\mathcal{A}_{k_{i}}\Omega_{m_{l}k_{j}%
}+\mathcal{A}_{m_{l}}\Omega_{k_{j}k_{i}}).
\end{equation}
However, this expression for the DMI coefficient is only locally defined
because of the gauge dependence of the Chern-Simons form \cite{Xiao2020AIOM}.

\begin{figure}[ptb]
\includegraphics[width=0.9\columnwidth]{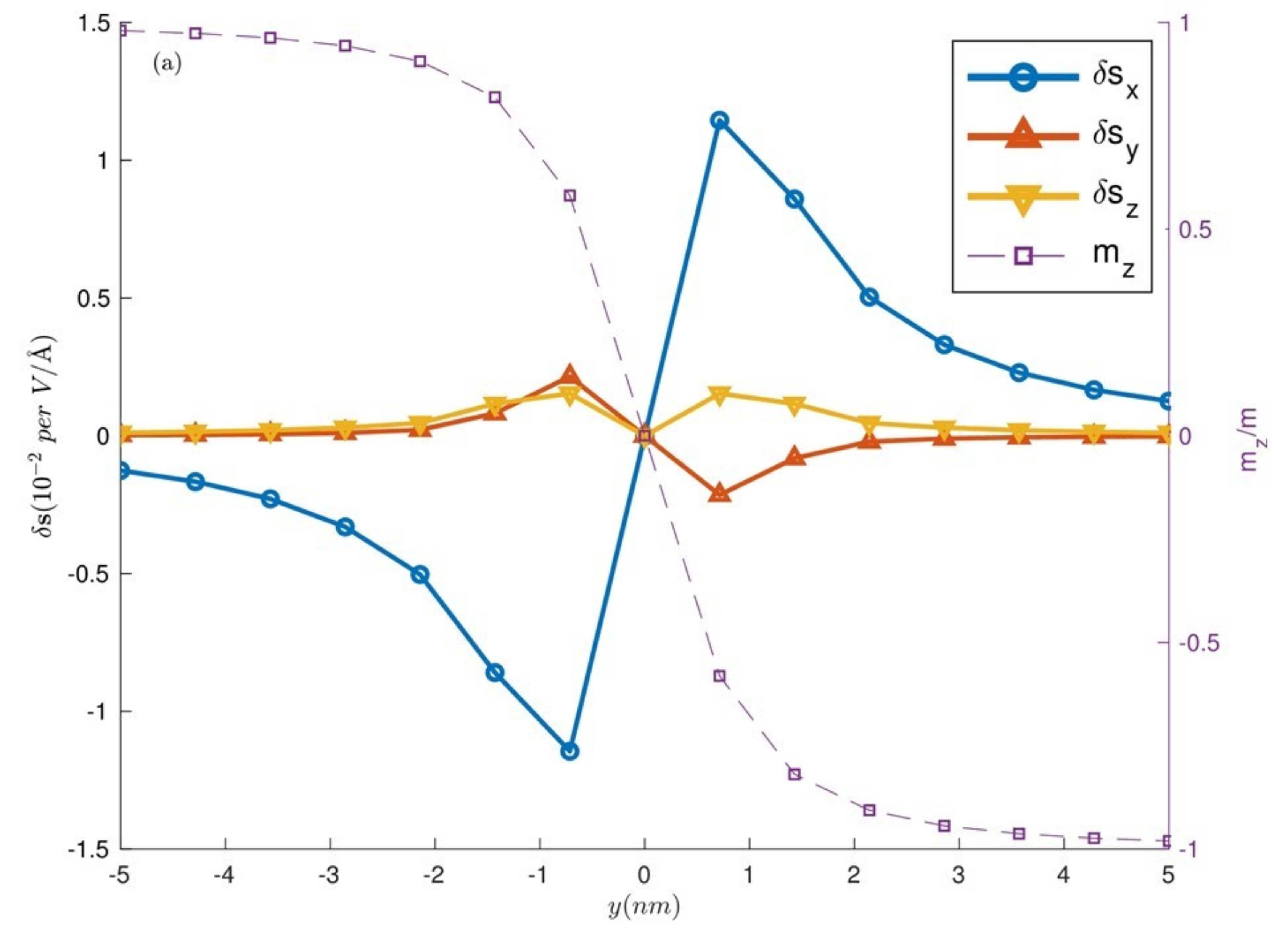}\caption{Spin generation due
to the electrical generalized force in a model of chiral Neel wall of
ferromagnetic transition metal dichalcogenide monolayer.}%
\label{fig:Neel}%
\end{figure}

On the other hand, the electrical DMI force
$\boldsymbol{\mathcal{\boldsymbol{E}}}_{m}^{e}$ [Eq. ({\ref{Chern}})] is not
only gauge invariant and single valued but also well defined for Chern
insulators. In practice, such a force enters directly in determining the
static and dynamic behavior of the magnetic textures. For example, we show in
the following that the width of a chiral Neel wall may be tuned by such a
force, as would normally be anticipated from DMI effects
\cite{Karnad2018,Kato2019}. Specifically, we consider a chiral Neel wall with
easy axis in the $z$ direction in a model of the insulating transition metal
dichalcogenide monolayer materials with magnetic proximity effect, and show
that its width would be enhanced (decreased) when an electric field is applied
in the $x$\ ($-x$) direction. We employ the model Hamiltonian $\hat{h}=\hat
{h}_{0}+J\hat{\boldsymbol{\sigma}}\cdot\boldsymbol{m}$, where $\hat{h}_{0}$ is
a six-band tight-binding Hamiltonian suitable for the low-energy physics in
monolayers of AB$_{2}$ (A = Mo, W; B = S, Se, Te), as was detailed in Ref.
\cite{Liu2013}. Consider a right-handed up-down Neel-type wall with easy axis
in the $z$ direction, the induced spin is plotted in Fig.~\ref{fig:Neel} under
an electric field in the positive $x$ direction. With the lowest two bands
filled, the first Chern form contribution vanishes. The dominant component,
$\delta\boldsymbol{s}_{x}$, is antisymmetric on the two sides of the domain
wall center. Thus the torque exerted on magnetization $\delta\boldsymbol{\tau
}=\delta\boldsymbol{s}\times\boldsymbol{m}$ is in the positive $y$ direction
on both sides, increasing the width of the domain wall. Apparently, when the
electric field is reversed, the width of the domain wall is
decreased.

\section{Conclusion}

In conclusion, we have studied nondissipative electric driving of
magnetization motion in uniform and nonuniform magnetic insulators due to
nontrivial topologies of occupied Bloch states in the combined space of
crystal momentum and magnetization configuration. The resultant
nonconservative electrical generalized force is capable of supporting
sustained magnetization motion even in the presence of Gilbert damping. A
minimal model has been exploited to show explicitly the limit-cycle behavior
of magnetic evolution. For magnetic textures, there is an additional
nonconservative and nondissipative electrical generalized force, related to a
Chern-Simons DMI for non-Chern insulators in the presence of an electric field.

\begin{acknowledgments}
We thank Hua Chen, Peng Yan, Yunshan Cao, Liang Dong and Tianlei Chai for useful discussions.
This work is supported by NSF (EFMA-1641101) and Welch Foundation (F-1255).
\end{acknowledgments}


\begin{thebibliography}{99}
\bibitem {Xiao2020AIOM}Under a gauge transformation of the wave functions, the
Berry connections acquire additional terms given by the gradients of a phase
field in the combined $\boldsymbol{k}$ and $\boldsymbol{m}$ space. This phase
is modulo $2\pi$ in the $\boldsymbol{k}$ space because of the periodic gauge,
and the $\boldsymbol{k}$ integral of these additional terms reduces to a
partial derivative of some scalar field in the $\boldsymbol{m}$ space.
Ordinarily, such a derivative term for the DMI coefficient is permissible,
because that would result in an immaterial total gradient in the DMI energy
density. When singularities occur as discussed in the example of the previous
section, this scalar field can be multi-valued over multiply connected regular
regions, rendering the DMI energy also multi-valued.
\end{thebibliography}
\end{document}